%% file: main.tex
\documentclass[sigconf,authorversion]{acmart}

\AtBeginDocument{%
  \providecommand\BibTeX{{%
    \normalfont B\kern-0.5em{\scshape i\kern-0.25em b}\kern-0.8em\TeX}}}

\copyrightyear{2024}
\acmYear{2024}
\setcopyright{rightsretained}
\acmConference[CHI '24]{Proceedings of the CHI Conference on Human Factors in Computing Systems}{May 11--16, 2024}{Honolulu, HI, USA}
\acmBooktitle{Proceedings of the CHI Conference on Human Factors in Computing Systems (CHI '24), May 11--16, 2024, Honolulu, HI, USA}
\acmDOI{10.1145/3613904.3642542}
\acmISBN{979-8-4007-0330-0/24/05}

\usepackage{color, colortbl}
\usepackage{multirow}
\usepackage{subfig}
\usepackage{longtable}
\usepackage{hyperref}
\usepackage[font=small,labelfont=bf]{caption}
\usepackage{wasysym}
\usepackage[nolist,nohyperlinks, printonlyused, withpage]{acronym}
\usepackage{enumitem}
\usepackage{siunitx}

\newcommand{\themestyle}[1]{%
    {\textit{#1}}%
}

\begin{document}



\title[Ethical Caveats of Conversational User Interfaces]{Listening to the Voices: Describing Ethical Caveats of Conversational User Interfaces According to Experts and Frequent Users}

\author{Thomas Mildner}
\orcid{0000-0002-1712-0741}
\affiliation{%
  \institution{University of Bremen}
  \city{Bremen}
  \country{Germany}
  \postcode{28215}
}

\author{Orla Cooney}
\orcid{0000-0003-3586-3503}
\affiliation{%
  \institution{University College Dublin}
  \city{Dublin}
  \country{Ireland}
  \postcode{}
}

\author{Anna-Maria Meck}
\orcid{0000-0002-2083-3929}
\affiliation{%
  \institution{LMU Munich}
  \city{Munich}
  \country{Germany}
  \postcode{}
}

\author{Marion Bartl}
\orcid{0000-0002-8893-4961}
\affiliation{%
  \institution{University College Dublin}
  \city{Dublin}
  \country{Ireland}
  \postcode{}
}


\author{Gian-Luca Savino}
\orcid{0000-0002-1233-234X}
\orcid{1234-5678-9012}
\affiliation{%
  \institution{University of St.Gallen}
  \city{St.Gallen}
  \country{Switzerland}
  \postcode{}
}

\author{Philip R. Doyle}
\orcid{0000-0002-2686-8962}
\affiliation{%
  \institution{University College Dublin}
  \city{Dublin}
  \country{Ireland}
  \postcode{}
}

\author{Diego Garaialde}
\orcid{0000-0002-6034-2761}
\affiliation{%
  \institution{University College Dublin}
  \city{Dublin}
  \country{Ireland}
  \postcode{}
}

\author{Leigh Clark}
\orcid{0000-0002-9237-1057}
\affiliation{%
  \institution{Bold Insight UK}
  \city{London}
  \country{Great Britain}
  \postcode{}
}

\author{John Sloan}
\orcid{0000-0001-6105-6110}
\affiliation{%
  \institution{Trinity College Dublin}
  \city{Dublin}
  \country{Ireland}
  \postcode{}
}

\author{Nina Wenig}
\orcid{0009-0007-5660-5446}
\affiliation{%
  \institution{University of Bremen}
  \city{Bremen}
  \country{Germany}
  \postcode{}
}

\author{Rainer Malaka}
\orcid{0000-0001-6463-4828}
\affiliation{%
  \institution{University of Bremen}
  \city{Bremen}
  \country{Germany}
  \postcode{}
}

\author{Jasmin Niess}
\orcid{0000-0003-3529-0653}
\affiliation{%
  \institution{University of Oslo}
  \city{Oslo}
  \country{Norway}
  \postcode{}
}

\renewcommand{\shortauthors}{Mildner, et al.}

\begin{acronym}
\acro{HCI}{Human-Computer Interaction}
\end{acronym}

\begin{abstract}
Advances in natural language processing and understanding have led to a rapid growth in the popularity of conversational user interfaces (CUIs). While CUIs introduce novel benefits, they also yield risks that may exploit people's trust. Although research looking at unethical design deployed through graphical user interfaces (GUIs) established a thorough 
understanding
of so-called dark patterns,
there is a need to continue this discourse within the CUI community to understand potentially problematic interactions.
Addressing this gap, we interviewed 27 participants from three cohorts: researchers, practitioners, and frequent users of CUIs. Applying thematic analysis, we construct five themes reflecting each cohort's insights about ethical design challenges and introduce the CUI Expectation Cycle, bridging system capabilities and user expectations while considering each theme's ethical caveats. This research aims to inform future development of CUIs to consider ethical constraints while adopting a human-centred approach.
\end{abstract}

\ccsdesc[300]{Human-centered computing~Empirical studies in HCI}
\ccsdesc[500]{Human-centered computing~Interaction design theory, concepts and paradigms}
\ccsdesc[300]{Social and professional topics~User characteristics}
\ccsdesc[300]{Human-centered computing~Ubiquitous computing}

\keywords{CUI, conversational user interfaces, conversational agents, voice agents, chatbots, thematic analysis, ethical design, deceptive design patterns, dark patterns}


\maketitle


\section{Introduction}
Conversational systems have been around since the mid-20th century~\cite{weizenbaum_elizacomputer_1966, colby_1981}. However, recent technological advances in natural language processing and understanding have led to conversational user interface (CUI) interactions becoming common in many people's daily lives~\cite{jung_understanding_2020}. Today, CUIs come in two key forms: text-based chatbots, which are often used in commercial settings to fulfil customer service roles (e.g., IBM Watson~\footnote{\href{https://www.ibm.com/watson}{https://www.ibm.com/watson}}, Amazon Lex~\footnote{\href{https://aws.amazon.com/lex/}{https://aws.amazon.com/lex/}}), and voice assistants, which are commonly integrated into smartphones and smart home devices (e.g., Apple's Siri~\footnote{\href{https://www.apple.com/siri/}{https://www.apple.com/siri/}}, Amazon Alexa~\footnote{\href{https://alexa.amazon.com/}{https://alexa.amazon.com/}}, Google Assistant~\footnote{\href{https://assistant.google.com/}{https://assistant.google.com/}}). Following the claims of CUI developers, the technology promises novel ways to interact with service providers efficiently, intuitively, and seamlessly~\cite{starke_unifying_2022,reicherts_its_2022}. The rapid growth of these technologies is mirrored by increased interest among researchers within the human-computer interaction (HCI) community~\cite{sin_cui_2023,porcheron_cui_2020,doyle_cui_2020}. 

Yet, many users experience frustration when engaging with CUIs~\cite{cowan_what_2017} that negatively impacts their overall experience. According to the literature, frustrations stem from exaggerated expectations among users regarding communicative competence that current CUIs cannot live up to. For instance, anthropomorphic CUI design can lead to unrealistic expectations of near-human-level performance when interacting with such devices~\cite{doyle_mapping_2019}. However, what users generally experience is a call-and-response type interaction rather than a free-flowing conversation. While recent efforts in generative artificial intelligence (AI) mitigate some of these shortcomings of traditional CUI interactions, recent work shows similar limitations remaining in large language models (LLM)~\cite{jo_understanding_2023}. In a similar vein, privacy and trust-related issues are said to limit the scope of tasks that users are willing to perform with a system~\cite{luger_like_2016, doyle_mapping_2019}. The lack of transparency around how personal data is recorded and used can also become an obstacle for users in taking full advantage of CUI capabilities, again limiting the contexts in which people are willing to use them~\cite{cowan_what_2017, porcheron_voice_2018}. 
To mitigate these issues, recent work offers domain-specific design heuristics~\cite{langevin_heuristic_2021} or frameworks to guide user engagement~\cite{yeh_how_2022}. Still, further guidelines are needed as the technology advances from its infancy. Suggesting a lack of sufficient guidelines, practitioners adopt graphical user interfaces (GUIs) best practices to account for the unique affordances that CUIs present~\cite{holmes_usability_2019, langevin_heuristic_2021, clarc_speech_2019}. However, we ought to be wary of the ethical implications as recent work in HCI voices serious concerns regarding unethical practices in contemporary GUIs -- providing a taxonomy of deceptive design patterns often referred to as ``dark patterns'', which inhibit users' decision-making to benefit service providers~\cite{brignull_deceptive_2022}. Although malicious incentives prevail in CUIs~\cite{owens_deceptive_2022, conca_present_2023}, we currently have a limited understanding of how unethical design manifests in the context of CUIs and which ethical caveats require consideration.
We, therefore, see an opportunity to proactively address concerns akin to recently addressed deceptive design issues that required legislative and regulatory actions~\cite{gdpr_2016,ccpa_2018,eu_dsa_2022} to protect users in GUI contexts. 

To that end, we aim to gain insights regarding ethical caveats for CUI design. Here, the dark pattern discourse supports this endeavour by providing a novel angle to understand unethical design in problematic CUI interactions. 
In the context of this work, we refer to \textit{ethical caveats} as considerations practitioners need to make to avoid adopting design strategies or features that undermine users' autonomy or deceive them into making choices that are not necessarily in their best interests. To that end, this work explores ethical concerns around the design of current CUI systems, as well as potential issues that may arise in the future, among three specific cohorts: (1) researchers who focus their work on CUIs, (2) practitioners who develop CUIs, 
and (3) frequent users who engage with CUIs at least once a week.
To our knowledge, this work presents the first to consider multiple perspectives to assess CUIs based on ethical caveats. 
In total, we interviewed 27 people to address the following research question:


\begin{itemize}
    \item [\textbf{RQ:}] Which ethical caveats should be considered when designing CUI interactions, and how should they be addressed?
\end{itemize}

In answering this research question, this work has two main contributions. Firstly, five themes outline ethical caveats in CUI design: 
\themestyle{Building Trust and Guarding Privacy}, \themestyle{Guiding Through Interactions}, \themestyle{Human-like Harmony}, \themestyle{Inclusivity and Diversity}, and \themestyle{Setting Expectations}.
Secondly, we introduce the CUI Expectation Cycle, a framework promoting ethical design considerations by incorporating our five themes. This framework responds to repeated demands for design guidelines to mitigate problematic interactions and negative user experiences expressed by practitioners while addressing users' concerns and expectations.

\section{Related Work}
The related work of this paper encompasses the current development of ethical awareness in HCI and CUI research. The section begins with recent work from the CUI community, addressing design and technology limitations and showcasing a need for explicit guidelines for practitioners. We will then briefly outline work addressing unethical design practices (a.k.a. dark patterns), currently most widely discussed in GUI research. The lack of such work in CUI research emphasises a need to consider similar issues when developing chatbots and voice assistants and motivates our research. Lastly, we will review recent advances in developing CUI-related guidelines and best practices.

\subsection{Design and Technology-based Limitations}
Due to the distinct affordances they present, CUIs have to overcome design challenges that cannot simply be borrowed from other fields~\cite{murad_2020}. In particular, the use of language, uttered or typed, as a primary input for CUIs poses problems regarding the ease with which users can explore their functionalities~\cite{branham_blind_users_2019,chen_technical_2018,wagner_alexa_2019} or restricting the assessment of a system's capabilities and boundaries~\cite{luger_like_2016, jain_2018, feine_2019}. This is affirmed by users' inability to recall the exact command needed for a response from their device~\cite{owens_deceptive_2022, corbett_what_2016}, limiting interactions to memorised prompts.

Further difficulties are linked to expectations users have toward CUIs. These expectations are steep as it is proposed that users can talk to their devices intuitively via a principal way of communication: language. However, in daily use, CUIs often fall short of their users’ expectations~\cite{luger_like_2016, doyle_mapping_2019, cowan_what_2017, Völkel_perfect_2021}. The degree of dissatisfaction this encourages among users has even led some experts in the field to suggest practitioners should rethink the ``Conversational'' part of ``Conversational User Interfaces'' altogether~\cite{Reeves_conversation_2019}. 
High user expectations also stem from anthropomorphic design features that encourage users to see CUIs as ``social actors'', a term coined by Nass and Brave in their Computers are Social Actors (CASA) paradigm~\cite{nass_machines_2000}. Anthropomorphic design has measurable effects through users aligning their speech patterns when talking to CUIs, as explored by Cowan et al.~\cite{cowan_voice_2015}. Although anthropomorphic design can serve beneficial purposes such as increased technology acceptance~\cite{cowan_voice_2015, kontogiorgos_effects_2019, seymour_exploring_2021, wagner_alexa_2019} and can influence the user experience positively~\cite{wagner_alexa_2019}, there are several usability ~\cite{kontogiorgos_effects_2019} and ethical problems ~\cite{bender_dangers_2021, garg_2020} associated with overusing anthropomorphic design features. Lacey and Cauwell~\cite{lacey_cuteness_2019}, for instance, studied dark patterns in connection to social robots, which could exploit ``cuteness'' attributes to coerce users' decisions and are only possible through anthropomorphic design. 

While Seymour and Van Kleek~\cite{seymour_exploring_2021} found a link between relationship development and anthropomorphic CUI design, Kontogiorgos et al.~\cite{kontogiorgos_effects_2019} voice a note of caution: although an anthropomorphic social agent led to a higher degree of engagement and sociability compared to a smart speaker without human-like traits, they also found that task-completion time with an embodied anthropomorphic agent was 10\% higher compared to a disembodied smart speaker. They link this to the anthropomorphic agent being perceived as a ``socially present partner in conversation''~\cite[p 138]{kontogiorgos_effects_2019}. In a similar vein, Lee et al.~\cite{lee2020} note a preference in users towards physically present CUIs over invisible ones based on a user study describing their mental model when interacting with CUIs. 
In recent work, Dubiel et al.~\cite{dubiel_conversational_2022} spotlight a connection between trust and anthropomorphic traits of CUIs. Realising ethical caveats, the authors propose four design strategies to calibrate trust while avoiding anthropomorphic features to enhance user engagement.
Hence, whilst anthropomorphic agents often seem to be preferred, they can lead to increased expectations and inadequate attributions of capabilities. This makes the design and implementation of human-like cues a delicate balancing act. Including human traits in CUI design can foster engagement and acceptance. However, concurrently, potential negative consequences for usability if users overestimate their CUI’s abilities need to be considered.

Problems also arise from more general technical difficulties associated with datasets used to train LLMs that underpin CUI capabilities. Specifically, LLMs are said to contain and reproduce highly biased worldviews, leading them to ``overrepresent hegemonic viewpoints''~\cite[p 610]{bender_dangers_2021}. A lack of data representing marginalised social groups increases their marginalisation even further. This is the case, for instance, with older adults, users with speech impairments or pronounced idiosyncrasies, or speakers with a strong colloquial dialect, vernacular, and/or accent for whom speech recognition performs poorer than for standard language speakers~\cite{koenecke_disparities_2020, mengesha_asr_2021,balasuriya_disabilities_2018, pradhan_disabilities_2018,harrington_its_2022}. As a result, diverse cultural experiences and identities are under-catered, leading to ``othering'' 
referring to the alienation of certain (social) groups resulting in marginalisation or exclusion in social contexts~\cite{mengesha_asr_2021}.
Where contemporary and commercially available CUIs initially promise natural language understanding and natural interactions, this promise only holds for a certain group of users. For example, African-American Vernacular English (AAVE) speakers need to adapt their speaking style and/or apply less natural speech patterns, like code-switching, to cater for a CUI's limited capabilities~\cite{harrington_its_2022}. These technological and design limitations hint at a lack of user-centred approaches at the end-user's expense. This is highlighted in work done by Blair and Abdullah~\cite{blair2020}, who identify the challenges of deaf and hard-of-hearing individuals using smart assistants in daily contexts. If marginalised groups are neglected in development processes, many will be unable to use the resulting systems~\cite{blair2020,harrington_its_2022}, posing unethical consequences~\cite{henrich_weird_2010}. This work considers different perspectives and, in part, investigates how marginalised groups can be better addressed in the design phases of CUIs.

\subsection{Summary of Dark Patterns}
Research on dark patterns has illustrated a wide range of applications where problematic design occurs~\cite{mathur_what_2021,gray_2018_dark,mildner_about_2023,zagal_dark_2013} while demonstrating a limited focus in the scope of CUIs~\cite{owens_deceptive_2022,conca_present_2023}. 
Adjacent literature guides our work to understand design areas that require further attention from practitioners to avoid unethical design in CUIs.
After Brignull coined the term ``dark patterns'' in 2010~\cite{brignull_deceptive_2022} and initialised a first set of twelve dark patterns, numerous researchers have set out to describe unethical practices in GUI interfaces. In 2022, however, Brignull promoted the term ``deceptive designs'' instead of dark patterns to address the risk of the term being racially misappropriated~\cite{brignull_deceptive_2022}. Recently, the ACM Diversity, Equity, and Inclusion Council~\cite{acm_words_2023} added the term to their list of controversial terminologies. However, critique against the term ``deceptive design'' has been voiced as it is deemed too vague to describe the scope and precision of its predecessor, lacking reference to pattern language~\cite{alexander_pattern_1977} while linking ``dark'' not to malicious intent but something that is hidden~\cite{obi_lets_2022}. 
Another argument is that such interfaces do not only deceive but also obstruct, coerce or manipulate users~\cite{mathur_what_2021, obi_lets_2022}. We acknowledge that, at the time of writing this paper, no perfect term exists to convey all unethical and problematic issues. As the community seeks a better term~\cite{gray_dark_23}, we opted to use the term ``dark patterns'' in this work to maintain consistency and continuity with previous research. 

Today, the dark patterns terrain has widened throughout numerous domains, including, but not limited to, mobile applications~\cite{digeronimo2020}, e-commerce~\cite{mathur2019, gray_2018_dark}, and social media~\cite{mildner_ethical_2021, mildner_about_2023, mathur_what_2021, gunawan_comparative_2021, schaffner_understanding_2022}. Collectively, dark pattern research has produced a taxonomy of individual design strategies (for an overview, we refer to Gray et al.'s ontology~\cite{gray_ontology_2023}). In an attempt to capture the diverse nature of dark patterns, Mathur et al.~\cite{mathur2019} describe five characteristics: \textit{asymmetry}, \textit{covert}, \textit{deceptive}, \textit{hides information}, and \textit{restrictive} (see Appendix~\ref{app:interview_questionnaire}, Table~\ref{tab:mathur2019} for a more detailed overview of the characteristics).
The authors ground their work on prior dark pattern research and their study based on a large-scale investigation of over 11,000 shopping websites. Each characteristic is introduced as a dimension where dark patterns manifest, while a single dark pattern can contain aspects of multiple characteristics. Moreover, characteristics are described through distinct mechanisms that restrict informed decision-making (e.g. user interfaces promote specific choices over others or hide relevant information from the user).
Our work utilises these characteristics during the interviews to learn about interviewees' views on potential unethical designs in CUIs.

Despite this growing field of research, most work focuses on GUI artefacts, and only limited work considers dark patterns in the context of CUIs. Surveying CUI users on twelve scenarios involving voice-based systems, Owen et al.~\cite{owens_deceptive_2022} establish a ground for future work by highlighting the importance of considering dark patterns in this environment, which finds further support by the provocation by Mildner et al.~\cite{mildner_rules_2022} from the same year. 
The need for further research is also mirrored in a recent review by De Conca~\cite{conca_present_2023}, who showcases the presence of known dark patterns in speech-based CUI systems. Thereby, the work not only highlights potential differences of dark patterns in CUIs compared to their GUI counterparts. The work further concerns necessary regulatory actions, emphasising the need for more tailored considerations, as contemporary regulation mainly concerns GUI dark patterns.

Importantly, studies have repeatedly shown difficulty among users to recognise and identify dark patterns sufficiently~\cite{digeronimo2020}, or with low accuracy~\cite{bongard-blanchy_i_2021}, even if provided information about dark patterns~\cite{bongard-blanchy_i_2021, mildner_defending_2023}. Users' inability to safeguard themselves places them in a vulnerable and exploitable situation that requires particular design considerations. In this context, it would be beneficial for practitioners to be guided by ethically aligned guidelines, which could aid in the development of user-centred systems that potentially empower users to make informed decisions.


Setting out to protect end-users of technologies, we aim to increase our understanding in this field by consulting practitioners, researchers, and frequent users. Led by the momentum of this discourse, we aim to draw attention to ethical caveats in CUI design as the technology becomes more ubiquitous. As practitioners often work under diverse ethical constraints, leading to the unconscious or unwilling deployment of unethical design~\cite{gray_ethical_2019}, we aim to provide concepts to avoid implementing dark patterns to begin with.

\subsection{Guidelines \& Best-Practices}
Spanning work including Nielsen's ten usability heuristics~\cite{nielsen_heuristic_1990} and Friedman et al.'s Value Sensitive Design~\cite{friedman_value_2013}, HCI has produced a range of important frameworks, guidelines, and best practices to aid practitioner' efforts. Yet, often recorded frustration and negative experiences of CUI users indicate a lack of application or applicability of these aids in this domain. This issue has been addressed by various researchers~\cite{holmes_usability_2019, langevin_heuristic_2021, murad_2020}. Ghosh et al.~\cite{ghosh_assessing_2018} highlights a lack of accuracy for the system usability score~\cite{brooke1996sus, brooke_sus_2013} --- a prominent measure of subjective perceptions of usability of systems -- when used to evaluate CUIs. To address this problem, recent work investigates the possibility of transferring concepts known to work for GUIs and other interfaces toward the context of CUIs. Langevin et al.~\cite{langevin_heuristic_2021}, for instance, adapt Nielsen's heuristics within the context of CUI interaction. Similarly, Klein et al.~\cite{klein_ueq_2020} revise the widely used UEQ questionnaire~\cite{laugwitz_2008} and extend it with scales for response behaviour, response quality, and comprehensibility to mirror aspects important for the user experience of CUIs. Indeed, various reviews of CUI research note the lack of validated measures of user perception as an ongoing problem that calls into question the reliability of this body of work due to a lack of continuity in how concepts are defined~\cite{clarc_speech_2019, seaborn_voice_2022, kocabalil_measuring_2018}. To date, there has been only one validated scale of this nature available to CUI researchers, known as the partner modelling questionnaire (PMQ) ~\cite{doyle_what_2021}.

Still, guidelines tailored explicitly to CUIs are scarce. Additionally, they often exclude unique requirements of vulnerable groups who could especially benefit from interactions with hands-free and speech-based devices, as is the case for users with, for example, visual impairments~\cite{branham_blind_users_2019}. Through our established themes and our framework, our work aims to provide some mindful guidance for practitioners and researchers to utilise a user-centred approach and reach a variety of different user groups.

\section{Method}
This study aimed to gain insights regarding CUI-specific ethical caveats and identify ways unethical practices manifest within CUIs. We, therefore, conducted a total of 27 semi-structured interviews split between three groups. The first group includes researchers focusing on CUI-related topics (N=9). The second cohort comprises practitioners who work in industries developing CUI technologies (N=8), whilst the third encompasses frequent users of CUI systems (N=10). The interviews were conducted online via video conference tools, which allowed for the recruitment of participants with an international scope.

\subsection{Interview Protocol}
The interview consisted of two parts. The first part focused on participants' general experience with CUIs. The second part is based on and inspired by Mathur et al.'s dark pattern characteristics, promising interesting insights into unethical design strategies
by describing design choices and mechanisms that prohibit informed decision-making. 
As Mathur et al.'s definitions were constructed to convey similarities between certain groups of dark patterns, the original definitions were neither created for an interview context nor designed to account for CUI-based interaction. To address these limitations, we adapted the original questions to foster more relevant answers from our participants. While this enabled us to learn about participants' views on dark patterns in this context, we also added three questions targeting specific situations and demographics to gain a deeper understanding of circumstantial issues. The full interview protocol and the five dark pattern characteristics according to Mathur et al.~\cite{mathur2019} are included in the Appendix~\ref{app:interview_questionnaire}. During the interview, participants were conditionally prompted to think about text-based and voice-based systems, both concerning their actual experiences and hypothetical future scenarios they might envisage.

\subsection{Participants}
In total, 27 participants were recruited for interviews. Researcher and practitioner cohorts were recruited from the authors' collective professional network and word of mouth, enlisting academic and commercial researchers, designers, and developers whose work focuses on CUIs. The third group, frequent users, were recruited via the online platform \textit{Prolific}~\cite{prolific}. Recruitment criteria were used to ensure participants in this cohort were at least 18 years of age and used CUIs at least on a weekly basis, though no particular CUI system was stipulated. All participants were informed about the nature of the study, what participation involved, and their data rights before being asked to provide informed consent. 
Participants recruited via Prolific were rewarded with a \textsterling10 honorarium. This is in keeping with suggested hourly rates for ethical payments for research involving crowdworkers~\cite{lascau_monotasking_2019}. Researchers and practitioners participated voluntarily and were not furnished with an honorarium. Table~\ref{tab:participants} presents a full overview of all recruited participants.
The interviews lasted an average of 42:30 minutes ($SD=$ 11:29). Interviews with the researcher cohort took the longest (mean=48:00, $SD=$ 08:48), while interviews with those held practitioners were slightly shorter on average (mean=44:47, $SD=$ 13:50). The interviews conducted with frequent users were the shortest (mean=36:03, $ SD=$ 08:28).

\input{Tables/particiant-table}

\paragraph{Researchers} Eight academic researchers were recruited for our study (three female, four male, one preferred not to disclose their gender), four of whom were professors, one a postdoctoral researcher, and three who were PhD candidates. When conducting this study, the researchers' average age was 37.0 ($SD=13.0$), and their average years of experience researching  CUIs was 13.9 years ($SD=14.4$).

\paragraph{Practitioners} Nine practitioners participated in our interviews (five female, four male) with roles in conversation design (2), executive management (4), research (2), and consultation (2). Practitioners' average age was 47.0 years ($SD=11.0$), and they had an average of 15.3 years ($SD=10.3$) of experience working in this space. 

\paragraph{Frequent Users} Finally, we recruited ten individuals for the frequent users' cohort (five female, five male; mean age $= 31.7$ years, $SD = 11$). To qualify as frequent users in this study, participants of this cohort had to use CUIs at least once per week basis. However, we did not require participants of this group to have long-term experience as the technology is still relatively young. We also asked them about the kinds of devices they used most often. Seven stated that they mostly accessed CUIs through their smartphones, whilst the other three mostly used smart speakers. CUIs used included Amazon's Alexa, Apple's Siri, Google Assistant, Microsoft's Cortana, Samsung Bixby, and text-based chatbots from online services. Six said they used CUIs daily, while four stated using CUIs more than once a week. The highest level of education among this cohort was: undergraduate (4); secondary/vocational (3); postgraduate or higher (3). Participants' occupations span construction (1), customer service (1), front-end developer (1), media analyst (1), medical laboratory scientist (1), self-employed healthcare entrepreneur (1), students (2), and teacher (1).


\section{Thematic Analysis}
After completing all 27 interviews, we transcribed and prepared the recorded material for analysis, de-identifying any traceable or personal information. Data transcription was carried out by a professional UK-based service provider. 
Based on these data, we conducted a reflexive thematic analysis~\cite{braun_thematic_2019}, which was divided into four phases: familiarisation; code generation; construction of themes; and revising and refining themes. 

\subsection{Positionality}
Authors of this work lived, gained education, and worked in Central Europe most of their lives, with WEIRD (Western, Educated, Industrialised, Rich, and Democratic)~\cite{henrich_weird_2010} backgrounds. Their research backgrounds contain expertise in design, linguistics, computer science, and psychology, with scholarly work oriented toward social justice and well-being topics.
The interview transcripts were coded by four authors who have previously engaged in scholarly work in human-computer interaction, computer science, and psychology, each contributing more than three years of experience in their respective domains.
At the time of conducting the study, no personal or professional conflicts with CUI systems or their developers occurred among the authors. Participants of the researcher and practitioner cohorts were recruited through professional networks with some personal relations. Where an author had a rather close relationship with a participant, we ensured that another author would conduct the interview instead. Frequent users were recruited through Prolific~\cite{prolific}, where no previous relationships existed. 
Focused on ethical considerations in CUI design, this work is partly inspired by Mathur et al.'s~\cite{mathur2019} dark pattern characteristics to analyse problematic CUI interactions and propose counter-measures. To conclude, we acknowledge potential bias based on our cultural, academic, and personal backgrounds.

\subsection{Coding of the Transcripts and Identifying Themes}
For the first round of data analysis, we selected two interviews per participant group to generate an initial set of inductive codes. Following advice from Braun et al.~\cite{braun_thematic_2019}, two researchers coded these six interviews, discussed their strategies and results, and drafted an initial codebook encompassing 65 codes. Through axial coding, the codebook was then reduced to 46 codes.
All interview transcripts were then split among four authors. Each interview was thus coded by a single author following Braun et al.'s proposed methodology~\cite{braun_2006}. Although coding was carried out independently, authors did meet to discuss the procedure once before the coding, once after half of the interviews had been coded, and one last time after coding had been completed. These discussions ensured coding strategies were aligned between authors, including resolving any issues authors may have had applying specific codes~\cite{braun_thematic_2019} and discussing any potential new discoveries in the data. Due to the collaborative nature of this process, indices of inter-rater reliability~\cite{braun_thematic_2019} are not provided. 

Four authors then went on to identify themes for each cohort independently. Based on the codes, quotes, and annotations, each researcher applied affinity diagramming~\cite{blandford_qualitative_2016} to develop early iterations of these. During this process, these first themes were evaluated and refined by iteratively checking each for applicability to interview material. They were then discussed among researchers to establish agreement. Here, we followed Braun and Clarke's theme mapping~\cite{braun_2006} to collapse similar themes. If a theme was adapted, it was again revised against the whole dataset and other themes before being accepted. This last phase resulted in the construction of five themes and a corresponding ethical caveat for each.


\section{Findings}
\input{Tables/ethical-caveats}
In this section, we describe the constructed themes based on reflexive thematic analysis. We highlight these themes across each of the interview cohorts individually. Notably, a degree of overlap is to be expected as themes tend to be intertwined, often in a supportive fashion. In total, we synthesised five high-level themes summarised in Table~\ref{tab:theme-descriptions}, where each theme is listed next to a statement for a particular ethical caveat. Additionally, we formulated guiding questions that refer to the ethical caveat to support the design of CUI interactions. Before each theme is addressed in detail, we present a high-level overview of our findings. 
The outline for each theme follows the same structure: a description followed by the perspectives of each of the three cohorts interviewed (researchers, practitioners, and users).

\subsection{Building Trust and Guarding Privacy: Operating Extrinsic and Intrinsic Factors}
This theme spotlights extrinsic and intrinsic challenges that result in trust and privacy deficits in CUI interactions. Across cohorts, a reoccurring theme echoed fears of untrustworthy handling of personal data, kindled by prior negative experiences and the reputation of companies and practitioners. A call from researchers to increase transparency to bridge users' concerns was underlined by users’ mention of their own safeguarding strategies, which limit interactions to basic functionalities and do not require them to disclose private information. Practitioners acknowledged these problems but noted counterintuitive regulations and design limitations obscuring access to settings rooted in speech-based interactions.

\paragraph{Researcher}
Researchers interviewed were aware of trust-related problems of CUIs, suggesting a need for greater transparency,  ``\textit{it's really difficult for users to know how data travels across different platforms}'' (R4). Connecting this problem to the ``\textit{reputation of the companies behind each device}'' one participant said that ``\textit{there's very little transparency [...] about how that data is being used and processed}'' (R3). Another link is drawn to deceptions based on human-like features with which ``\textit{we are undermining the trust when we are projecting something which isn't real}'' (R5). Meanwhile, there are ``\textit{ethical issues with data sharing and things like that, but [users] care more about: `Can I get to what I need to get to fast enough?'}'' (R4). Looking at users' uncertainty from a technological lens, another researcher discussed how ``\textit{people didn't know [the CUI] was listening to them all the time [...]. How does it know when I say `Hey Alexa' or `Hey Google' [...]? Is it already listening to me?}'' (R2).

\paragraph{Practitioner}
Practitioners also reflected on privacy and trust issues. The current situation users are in when engaging with CUIs was described as ``\textit{completely wild west, talking to a black hole}'', because ``\textit{sometimes there is a lack of transparency, you just don't know what they're doing with your information and that's a problem for a lot of people''} (P8). CUI's present some inherent difficulties when users try to access their privacy settings: ``\textit{That info does seem to be kind of buried [...] you have to click in three or four different menus to get to the privacy stuff and I don't think it's explained}'' (P6). It is later reflected that users ``\textit{don't get any kind of introduction to the app itself or the onboarding is very light}'' (P6). 
Regarding a lack of technological literacy among users to understand and trust their CUIs, one practitioner asked the question: ``\textit{How do you prevent people from overhearing my voice and trying to imitate me, how does that thing know that it’s not really me? }''. They went on to provide an answer: ``\textit{There’s a simple explanation. It’s too smart, it knows the difference. The technical explanation is, it gets into how it analyses your voice}'' (P3).

\paragraph{Frequent User}
Being the affected cohort, users mentioned various concerns regarding their privacy and why they hesitate to trust CUIs. One participant was ``\textit{not fully comfortable disclosing [their] personal details even if it's a virtual assistant}'' (F5). Reflecting on potential trust connected to anthropomorphising CUIs, they later stated that ``\textit{it feels like you're speaking to a human and you're giving them your personal data}'' (F5). Talking about giving consent to things they never fully read, one participant said: ``\textit{I don't think I've read them all, I just think that I know what I'm accepting. So I don't know any of the consequences that I might have with it}'' (F3). 
Users are ``\textit{expected to know the risks to [their] privacy, to data, to enter into the real world. [Users] are expected to know it in advance.}'' (F7). Some worry about exposing critical information publicly, for instance regarding their financial accounts: ``\textit{It puts me in danger of whoever that's around that can hear how much money I have. Maybe I'm at risk of being robbed}'' (F2). Despite their concerns, the interviews outlined aspects of a privacy paradox. ``\textit{Today, there is no alternative [to commercial CUIs]. [...] It is bad, but I don't see [an] alternative}'' (F8).

\subsection{Guiding Through Interactions: Overcoming Knowledge Gaps}
This theme illuminates the importance of providing users with informed guidance about possible CUI interactions, as a lack of technological literacy and limited experience results in difficulties in accessing available features. While researchers noted that research findings need to be better aligned with industry development, practitioners echoed a desire for further guidelines.

\paragraph{Researcher} 
Researchers reflected on a lack of communication between stakeholders, as well as peculiarities of CUI interaction that lead to unique restrictions. A level of trust and limited collaboration was emphasised by researchers, who suggested work in academia seems to ``\textit{make very little impact on what big companies do. [...] If it is profitable, they're going to do it}'' (R4). Another researcher speculated that there was a disconnect between the communities that design the components required to build CUIs, and the community of HCI researchers who explore and develop theories that explain how users interact with CUIs: 
``\textit{What I see is a bunch of different technology CUIs just being demonstrated, and then in our particular space [design research] we’ve got toolkits, and the toolkits define what’s possible. [...] I’m struggling really to understand how come there’s this disconnect between these communities?}'' (R5). It was noted that the communication between practitioners and users is also problematic ``\textit{if users don't care and only developers do}'' (R4).

\paragraph{Practitioner}
Practitioners again referred to a lack of best practices guidelines that would help them create better products. ``\textit{[...] There’s very little that’s actually focused on conversational user interfaces... even to the point that we don’t have heuristics really. [...] To create these heuristics for CUIs, that hasn’t been done, so there’s this massive gap out there.}''(P6).When contrasted against comments by researchers, this further highlights the disconnect between these communities.
Practitioners also highlighted the lack of standardised guidelines for users when it comes to how best to interact with these kinds of systems. They suggest users are essentially left to rely on what they know from graphic counterparts: ``\textit{For voice stuff, what's missing, there is no standard, like you know when you go to a website, and there's that little hamburger menu, those three lines.}'' (P6). Instead, users have to compare their experiences with other, similar systems in the hopes that they operate the same way; ``\textit{[...] it's still fairly new and [...] design best practice is still kind of evolving. [...] Maybe if there was some more standardisation in how the experience works between different companies or between different interfaces, people will get more used to it}'' (P4).

\paragraph{Frequent User}
Frequent users echo the idea that the burden of figuring out the limitations of a speech agent falls largely on themselves. ``\textit{Most of the time you have to understand Google Assistant and what its limitations are when you ask it to do specific tasks }'' (F5). Similarly, users ``\textit{have to use the correct terms}'' and ``\textit{learn to adapt to [their] language set}'' (F7). In reference to using a CUI on their smartphone, one participant reflects on the limited information available to them: 
``\textit{You'd have to actually sit on your phone and actually test out everything to see what works and what doesn't}'' (F3), whilst another suggests they draw expectations from smartphone experiences because ``\textit{they are able [...] and that's what we want from our home devices}'' (F8).

\subsection{Human-like Harmony: Providing Authentic Anthropomorphism}
This theme describes the importance of implementing an appropriate amount of human-like features so that a system can support intuitive interaction without eliciting false beliefs that may leave users feeling deceived. 
Finding the right balance between humanness and technological limitations is challenged by the phenomena of anthropomorphism. That is the tendency to attribute human characteristics to non-human objects that most people engage into some degree~\cite{waytz_2010} from early childhood~\cite{piaget_childs_1997}. Anthropomorphous behaviour appears significantly heightened in dialogue with technological devices endowed with gendered voices and names~\cite{gong_2007}. Other influences that might encourage anthropomorphous behaviour in this context include: expectations for social affordances implied by representations of speech interfaces in media and advertising~\cite{murad_2020}; the fact that these systems conduct tasks typically carried out by humans using human language~\cite{nass_1994}; and that language use itself might be inherently social and agentic~\cite{fausey_2010, jia_2013}. When developing CUIs, practitioners should conscientiously navigate the amount of human-like features to avoid inadvertently manipulating our bias for anthropomorphic characteristics.

\paragraph{Researcher}
Researcher interviewees provided thought-provoking ideas to enhance transparency around the ontological nature of speech agents and demonstrated awareness of potential ethical concerns associated with endowing CUIs with human traits. Interviewees suggested CUIs should be designed in such a way ``\textit{that people are not fooled into thinking it’s a social entity when it’s just a machine}'' (R7), and that artificial voices should be designed to enable users to differentiate between humans and machines easily (R1, R5, and R8). Some even went as far as to state that ``\textit{From an ethical standpoint [a CUI] should announce it’s a machine}'' (R5). This would undoubtedly provide clarity for some users who experience confusion due to ``\textit{not knowing what social situation [they are] in}'' (R1). Overall, researchers agreed there should be increased transparency for users when engaging with CUI technologies.

\paragraph{Practitioner}
The need to increase transparency was also acknowledged by individuals from the practitioner cohort, who suggested CUIs ``\textit{should flag the fact that it is a machine talking}'' (P9), arguing for the introduction of a ``\textit{watermark [...] that indicate[s] that [the CUI] is fake}'' (P8). Although ``\textit{some level of humanisation}'' was thought to cause no harm, this could be context-dependent (P9). Practitioners felt the responsibility lies in ``\textit{how the bot identifies itself [...,] how human-like the conversation is}'' and ``\textit{the visual representation''} (P9). However, there is a risk for users who ``\textit{still got [...] attached to [CUIs]}'' (P2), even though the interaction was not particularly ``\textit{emotional or personal}''. In this regard, a counterargument mentioned that ``\textit{people are lonely and [...] there is a benefit to having a virtual companion in some ways}'' (P2). Another practitioner did not see any issues endowing CUIs with human-like characteristics, believing people can still easily distinguish between humans and a device that ``\textit{just speaks like a human}'' (P10).

\paragraph{Frequent User}
Many of the frequent users' observations echoed facets of the other cohorts' observations. Demonstrating the kind of bond some users are said to have with their devices, one participant admits that ``\textit{it really feels like you have your friend walking with you in your pocket}'' (F5) when referring to the virtual assistant built into their mobile phone. Similarly, another participant finds that ``\textit{especially voice assistants [...] let users believe that they are human or, to an extent, friendly}'' (F6). It was felt that a lack of awareness may lead users to ``\textit{interact [with CUIs] as [they] would with a human without knowing that it is not}'' (F1), with one participant recalling how elderly family members ``treat the machine like a human being'' and use it ``\textit{with courtesy}'' (F10). However, some also highlighted limitations when interacting with certain CUIs, particularly those implemented in customer service. Expecting a human to help with a problem, the limited responses and understanding of queries exhibited by some systems were said to create frustration and exacerbate a desire to ``talk to a person, a real person'' (F10).

\subsection{Inclusivity and Diversity: CUIs in the Wild}
This theme delves into the unexpected design and interaction challenges emerging when CUIs are introduced to diverse user groups with distinct characteristics and needs. Participants spotlighted several groups susceptible to poor design choices or technical limitations within CUI interactions. They emphasised a regression towards the mean, acknowledging how CUIs are designed with an ``average'' user in mind, leading to the ``othering''~\cite{mengesha_asr_2021} -- the marginalisation or exclusion of certain people  -- of individuals and groups that do not fit this profile. This encompasses users with stronger accents, colloquial dialects, second language users, people with deficits in speech or cognition, or users with lower technical literacy.

\paragraph{Researcher}
Researchers note that in designing CUIs, one has to be aware of varying user abilities and behaviours; ``\textit{So certainly you have users with diminished capacity that are going to be particularly vulnerable.}'' (R1). This was seen as particularly problematic around consent ``\textit{if you know that certain groups of users can’t give their informed consent, yet they’re still using certain technologies, then you’d have to be more careful. 
}'' (R4) 

\paragraph{Practitioner}
Practitioners also suggested, ``\textit{[T]he use of these systems for vulnerable groups is problematic.}'' (P1), whilst also identifying issues faced by minority groups: ``\textit{in terms of demographic, in terms of like accent, people who have a strong accent because the language that they're trying to interact with isn't their native language. Basically, anyone who doesn't fall into the kind of the centre of the bell curve, I expect, is more susceptible [to problematic CUI design]}'' (P4).

\paragraph{Frequent User}
Frequent users identified a number of groups they envision encountering issues with CUIs; ``\textit {people who are not technology literate or digitally literate, they might be vulnerable.}'' (F5), ``\textit{there are certain people who are very gullible to [...] information in general, anything that is said online they absolutely believe.}'' 
(F2). Frequent users also offered anecdotal examples of individuals that have been [other-ed] by CUIs, explaining that ``\textit{ a user would need to have a fairly good memory to be able to use [CUIs] effectively. I have a friend who is severely disabled, and she just cannot use it because she can’t remember the commands.}'' (F2). Another participant shared that ``\textit{ a friend of mine, her son is autistic [...] he has trouble with her Alexa, [...] sometimes he knows the answer but he’ll ask the question for information. He’ll expect it to conform exactly to what he knows, even using the correct words and he gets so frustrated when it doesn’t}'' (F7).

\subsection{Setting Expectations: Transparency to Mitigate Frustration}
This theme emphasises the pivotal role of transparency when designing CUI interactions to set realistic expectations. Unintended device reactions can result in disappointment and frustration in the user when a device appears more capable than it actually is. To deliver users an optimal experience, commercial incentives should be aligned with human-centred practices of HCI to avoid ethical concerns tied to deceptive and manipulative interactions. Instead, transparent and evident CUI capabilities should empower users to use the system autonomously and easily. 

\paragraph{Researcher}
The importance of setting the right expectations for what systems are capable of was noted among researchers for offering users a better experience that is more aligned with reality. Here it was stated that ``\textit{functionality expectations are not set right}'' (R2) for contemporary CUIs, with introductions and manuals either not being readily available or ignored by users. Additionally, ``\textit{people bring their previous experiences to an interaction when they start}'' (R7), suggesting users with prior knowledge have an advantage over those who are new to CUIs. If true, this highlights a lack of onboarding for novice users, with prior knowledge biasing experiences for frequent users. In this regard, ``\textit{lack of transparency [...] really makes it easy to misunderstand what the system does, or is capable of doing}'' (R2).

\paragraph{Practitioner}
Practitioners suggest a lack of best practices or guidelines for how and when to communicate a systems' features to users; ``\textit{getting across to the user what the system can and cannot do is one of the most difficult problems}'' (P2). Current commercial CUIs are limited to presenting information synchronously on request. As the same metaphors working in GUI environments do not translate readily to CUIs, ``\textit{discovery [of functionalities] is very difficult right now because there's no directory to help us find it}'' (P8).  While multi-modal chatbots can mitigate certain problems, voice assistants without integrated displays cannot rely on the same design philosophies. ``\textit{specifically for voice stuff, what's missing, there is no standard}'' (P6). 
It was further argued that ``\textit{when [CUIs] can only answer a fraction of questions it is inherently putting the burden on the user to rephrase their question over and over}'' (P6). 
This difficulty is increased ``\textit{if there are a lot of options, it's probably more of a challenge to make people aware of the choices that they have}'' (P4). In relation to human-like features, how CUIs ``\textit{are designed can make them seem more intelligent than they are}'' (P9), creating unrealistic expectations that will eventually cause frustration in users. 

\paragraph{Frequent User}
Multiple users expressed frustrations they experience when using their CUIs; if a CUI ``\textit{fails to give you what you want it to do, you end up having to do it yourself}'' (F5). An initial problem stems from raised expectations through advertisement of commercial products, where ``\textit{everything [...] runs smoothly [...] creat[ing] that expectation in your head as that's how it works''} (F1). Another participant explains that if ``\textit{you don't know what [the CUI] has heard [...], you have to go to the app}'' (F10) as the device is incapable of providing insights into why errors occur. Limiting their expectations, one user adapted their behaviour to ``\textit{only [use voice-based systems] for easy communication because [...] they are not set yet for more complicated things}'' (F8). Another user discussed the importance of good memory, which ``\textit{users need to have [...] to be able to use [CUIs] effectively} and that they ``\textit{need to have read the literature for the [CUI] to know the limits}'' (F7).

\section{Addressing Expectations: The CUI Expectation Cycle}
Based on our findings, we constructed the CUI Expectation Cycle (CEC), a framework that practitioners and researchers might use to understand and serve users' needs better. 
To this end, the CEC aligns user expectations with system capabilities. It aims to enable successful CUI interactions that meet the goals of its users, mitigating unethical design strategies and voiced frustrations of our participants in line with related work~\cite{harrington_its_2022,yeh_how_2022}. The CEC incorporates the themes identified in our qualitative analysis, including garnered ethical caveats. Further, it partly builds on existing frameworks -- Norman's \textit{Action Cycle}~\cite{norman_design_2013} and Oliver's Expectation Confirmation Theory~\cite{oliver_effect_1977, oliver_cognitive_1980} --  to inform about the variables that shape user expectations. 

\subsection{Constructing the CEC}
\begin{figure*}[!t]
    \centering
    \includegraphics[width=1\textwidth]{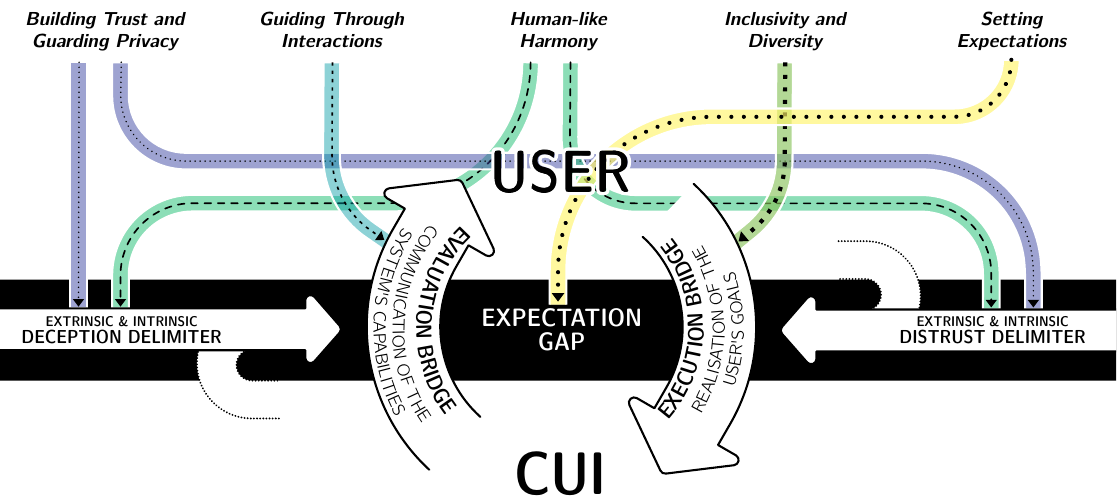}
    \caption{The CUI Expectation Cycle demonstrates how an expectation gap between user and system can be bridged by an evaluation and an execution bridge, inspired by Normans' two evaluation and execution gulfs~\cite{norman_design_2013}. However, deceptions can impact how users assess the system's capabilities, resulting in unrealistic expectations. Similarly, distrust limits users' faith in the system and its responses, influencing their execution of actions. Both delimiting factors can be of extrinsic as well as intrinsic nature. 
    The diagram further features the five themes and connects them to respective sections of the CUI Expectation Cycle.}
    \Description[Figure showing the CUI Expectation Cycle]{This figure presents the CUI Expectation Cycle. In the centre is the expectation gap, separating user and CUI systems through mismatched expectations. The expectation gap is connected to the "setting expectations" theme. Two arrows bridge this expectation gap. The first connects the user with the CUI and is called the execution bridge. It describes users' ability to realise their goals. The "inclusivity and diversity" theme is connected to this bridge. The second is the evaluation bridge, which returns from the CUI toward the user. It entails how well the CUI communicates its capabilities. Here, the "guiding through interactions" theme is further linked. The ideal state of this cycle is interrupted by two delimiting factors: Deception and distrust. Either can occur extrinsically and intrinsically, and both feature the "building trust and guarding privacy" and "human-like harmony" themes.}
    \label{fig:expectation-cycle}
\end{figure*}

Drawing from dark pattern literature, our interviews confirm the findings of existing CUI research while providing a detailed overview of the interplay of design challenges. While identifying our themes, we assessed their interrelationships and noticed a strong gap between users' expectations and CUI capabilities, as described by our participants. Focusing on this gap, we distilled the themes into a framework while consulting related work. 
To this end, the CEC (see Figure \ref{fig:expectation-cycle}) links core issues of CUI design with a particular focus on anthropomorphism~\cite{doyle_mapping_2019,doyle_what_2021}, deception~\cite{owens_deceptive_2022,conca_present_2023}, inclusivity~\cite{harrington_its_2022}, and trust~\cite{luger_like_2016,torre_trust_2018}. 
Moreover, the CEC builds on two established frameworks. On the one hand, it utilises Norman's Action Cycle~\cite{norman_design_2013} by adopting users' execution and evaluation gulfs; in the case of our framework, we refer to them as \textit{bridges} as they help to cross the Expectation Gap. We were also inspired by Oliver's Expectation Confirmation Theory~\cite{oliver_effect_1977, oliver_cognitive_1980}, a cognitive framework that describes customer expectations and subsequent (dis)satisfaction with purchased goods. Supported by these widely accepted frameworks and related literature, the CEC focuses on creating realistic expectations extended by carefully considering ethical caveats to avoid problematic design in the area of CUIs. Ideally, this framework enables the development of transparent and trustworthy CUI interactions, allowing satisfying and successful user engagement.


\subsection{The Elements of the CEC}
Figure~\ref{fig:expectation-cycle} illustrates the CEC and how the five themes relate to individual elements of the framework. Dividing user (at the top of the CEC) and CUI (at the bottom), the critical element of the CEC is the \textit{Expectation Gap} (at the centre of Figure~\ref{fig:expectation-cycle}). However, the \textit{Expectation Gap} can be overcome through two bridges that connect the CUI with the user (the \textit{Evaluation Bridge}, left of the centre) and vice versa (the \textit{Execution Bridge}, right of the centre).
In jeopardising these bridges, two delimiters result from problematic or unethical design decisions (the \textit{Deception Delimiter} on the outer left side and the \textit{Distrust Delimiter} on the outer right, each pointing toward the bridges).
Above the diagram are the five themes: \themestyle{Building Trust and Guarding Privacy}, \themestyle{Guiding Through Interactions}, \themestyle{Human-like Harmony}, \textit{Inclusivity and Diversity}, and \textit{Setting Expectations}. Each theme is connected to the relevant elements of the diagram through arrows with differing dotted lines and colours.
Informed by our themes, the elements of the CEC encompass the ethical caveats identified as considerations for the design of CUIs. Here, we describe each element more closely, beginning from the centre while highlighting elements and connected themes in \textit{italic} font.

The \textit{Expectation Gap} encompasses our \themestyle{Setting Expectations} theme, which reflects the different perspectives across cohorts regarding technical capabilities and design decisions. As these decisions anchor user goals, it is pertinent to design accessible interactions that address users' needs depending on a CUI's purpose and its situational context. For instance, frequent users reported frustration with the complexity in which CUIs had to be prompted, while practitioners noted how many features were left unused.
Consequently, reasonable bridges need to connect both sides. The CEC includes two such bridges (\textit{Execution Bridge} and \textit{Evaluation Bridge}) inspired by Norman's~\cite{norman_design_2013} gulfs of expectation and evaluation. 

To the left of the CEC's centre, the \textit{Evaluation Bridge} starts from the CUI and reaches toward the user. It thereby includes understandable and truthful communication of a CUI's capabilities to inform the user about possible interactions and set realistic expectations. This bridge is strongly connected to the \themestyle{Guiding Through Interactions} theme, expressing a need to empower users to fully understand possible interactions and their consequences.

The \textit{Execution Bridge}, right of the CEC's centre, starts at the user's side and links it to the CUI. The bridge describes how expectations are shaped into goals a user can realise. However, a variety of obstacles can hinder users from successfully executing interactions. In this regard, the \themestyle{Inclusivity and Diversity} theme addresses the importance of accommodating users' individual characteristics in the design of CUIs.

In a well-designed system, both bridges support users in making realistic assumptions about a system and using it without frustration. In a sub-optimal system, users may encounter negative experiences and designs lacking ethical considerations, which could adversely affect their well-being, as dark pattern literature indicates~\cite{mathur_what_2021, mildner_about_2023}.
The CEC includes two delimiters fostering the negative influences of the two bridges. The \textit{Deception Delimiter} (on the diagram's outer left side and pointing to the \textit{Evaluation Bridge}) has a detrimental effect on the \textit{Evaluation Bridge}, resulting in unrealistic expectations. Similarly, the \textit{Distrust Delimiter} (on the diagram's outer right side and pointing toward the \textit{Execution Bridge}) adversely affects the \textit{Execution Bridge} by decreasing users' faith in the system and its responses. The two delimiters are informed by our \themestyle{Building Trust and Guarding Privacy} and \themestyle{Human-like Harmony} themes as they advocate transparent and authentic interactions. The different perspectives cast across our participants highlighted how each delimiter can be intrinsic and extrinsic in nature. To illustrate, previous individual experiences impact how a user engages with a CUI (i.e. intrinsic in nature), while design decisions influence how a CUI is perceived (i.e. extrinsic in nature). Importantly, either delimiter can be the source of unethical design practices commonly found in dark patterns.


\subsection{Using the CUI Expectation Cycle}
Although previous research has repeatedly addressed a gap between users' expectations and CUI devices~\cite{luger_like_2016, doyle_mapping_2019, langevin_heuristic_2021}, our participants, especially practitioners, requested design guidelines to overcome related issues. The CEC responds to these demands through a user-centred approach, embedding ethical caveats to avoid problematic designs. While the CEC could be used as a standalone framework, we believe it is best utilised next to contemporary efforts in CUI research to provide additional design considerations and constraints as alternatives to GUI-related best practices.
To support practitioners' work, we created questions derived from the ethical caveats that allow tracing of related limitations (see Table~\ref{tab:theme-descriptions}). Following the CEC (Figure~\ref{fig:expectation-cycle}), practitioners can consult each theme in the corresponding design areas to ensure that ethical caveats are respected.

Generally, CUIs promise benefits for people with diverse abilities, like visual impairments~\cite{branham_blind_users_2019}. However, studies demonstrating difficulties among marginalised demographics~\cite{harrington_its_2022} suggest that these benefits are not exhausted, addressed in our \themestyle{Inclusivity and Diversity} theme. CUIs, such as Amazon Alexa, often require additional GUI input to control device settings or personal data, restricting access. To cater to diverse user groups, practitioners could consider including alternative interactions that allow simple execution of planned goals. Asking our guiding question -- ``Does the system/interaction cater towards users with diverse needs, potentially through alternative interactions where otherwise inaccessible'' -- would remind practitioners about the importance of accessibility in their systems. 

Taking the \themestyle{Human-like Harmony} theme, for example, anthropomorphism can impact the deception and distrust delimiters. To mitigate deriving consequences, chatbot systems should clarify if no human is in the loop and state the scope of possible prompts. Design borrowing from other chat interfaces can cause additional confusion but could easily be avoided through a distinctive interface design. Similarly, anthropomorphic voice-based CUIs may confuse some users, particularly those with little technical literacy, leveraging their expectations. Appropriate voice design could clarify a CUI's artificial nature, clarifying related misconceptions. Asking our guiding question -- ``Does the system/interaction clarify the presence of anthropomorphic features to avoid misconceptions and unrealistic expectations?'' -- helps to unravel otherwise deceiving design choices.

In line with prior work~\cite{langevin_heuristic_2021}, our findings demonstrate that simple adoption of GUI best practices do not readily translate to CUI interactions as heuristics differ. While GUIs utilise graphical icons, buttons, or links~\cite{owens_deceptive_2022}, with some exceptions, CUIs' general input lie in uttered or typed-in commands. Though CUIs, including a GUI, can use these interactions, voice-based devices often require prompts to be spoken only. In most current CUIs, however, interactions require users to remember the correct prompts for the expected response, as recalled by some participants. Consequently, practitioners should account for human errors, support them better during interactions, and offer suitable alternatives. Adequate onboarding of the user is a crucial step to lower barriers and set expectations. While the CEC cannot account for malintent, we envision following its recommendations facilitates a better user experience and helps mitigate risks for users in consideration of our proposed ethical caveats. 

\subsection{Retrospective Application of the CEC}
To assess the guiding utility of the CEC, we apply our framework retrospectively on a selection of four high-quality, contemporary papers relevant to CUI research while sharing topical overlaps with our contribution. Papers were chosen based on their research quality and rigour from esteemed conferences. Consequently, we apply the CEC to these papers' findings to illustrate its relevance while demonstrating how it can be used.


In their work, Yeh et al. \cite{yeh_how_2022} discuss common pitfalls of guiding users effectively in task-oriented chatbot interactions. A key finding of their study describes how lack of transparency segues into frustration. In addressing the complexity of making possible interactions evident to the user, the authors describe guidance strategies that inform users about a chatbot's capabilities adequately. In line with the CEC, the authors unravel the need to align expectations (\textit{Expectation Gap}) between user and system through guidance as featured in our \themestyle{Guiding Through Interactions} theme. Incorporated in the \textit{Evaluation Bridge}, our \textit{Guiding Through Interactions} theme helps overcome these difficulties.

Following the potential advantages of LLM-supported CUIs, Jo et al.~\cite{jo_understanding_2023} studied users' experience with a chatbot as an empathetic conversation partner in care environments. However, users reported a lack of emotional support, which they linked to limited personalisation and missing health history. Moreover, particular worries accompanied users' experience, for example, a fear that personalised and empathetic CUIs may reduce peoples' desire to engage in social activities. The authors identified problems and foreshadowed solutions akin to our \themestyle{Inclusivity and Diversity} theme and reflected by our Deceptive and Trust Delimiters. As the \themestyle{Setting Expectations} theme addresses requirements to bridge an \textit{Expectation Gap}, our \themestyle{Building Trust and Guarding Privacy} and \themestyle{Human-like Harmony} themes could answer user's worries by making the system more transparent and trustworthy, increasing their experience.


Voice-based systems introduce additional layers to the interaction that require consideration. Language and speech barriers become obstacles for many marginalised groups that require special attention. Based on a study run in the U.S.A., Harrington et al.~\cite{harrington_its_2022} showed how Black adults from lower-income households struggle when using voice-based CUIs. Participants mentioned falling back to code-switching, mentioning the fear of being misunderstood.
Additionally, the researchers captured distrust toward Google Home in the health-related context of their study.
These trust issues were amplified by a lack of knowledge and technological literacy, hindering users from experiencing the CUI's full potential. As a source for these problems, the authors identify a host in the ignorance of developers who neglect the cultural identity of their users.
Providing practitioners with further insights, our \themestyle{Inclusivity and Diversity} theme spotlights the importance of catering to diverse users, as ignorance and negligence can swiftly lead to unethical implications. As Harrington et al. describe, CUIs should foster trust and transparency (i.g. avoid \textit{Deception} and \textit{Distrust Delimiters}). 
Their study highlights the importance of careful and inclusive research design, given the WEIRD (western, educated, industrialized, rich, and democratic)~\cite{henrich_weird_2010} context in which many contemporary studies in HCI are conducted. 
It further demonstrates that some issues cannot be mitigated through LLMs alone.
 
Generally, LLMs open promising avenues for CUIs whether deployed through text or voice-based systems. Exploring these advantages to make mobile GUIs more accessible through an assistive CUI, Wang et al.~\cite{wang_enabling_2023} describe prompting techniques that allow users to engage with interfaces through speech, particularly helping marginalised and vulnerable groups to access otherwise unattainable content. Their work is an excellent example of how the CEC's \textit{Expectation Gap} can be bridged by showcasing how conversations could be customised around special needs, thereby increasing the inclusivity of this technology in the future in line with our \themestyle{Inclusivity and Diversity} theme.

Collectively, these works illustrate relevant design challenges that the CEC emphasises. The addition of ethical caveats, elevating user experience, finds support as our \themestyle{Inclusivity and Diversity} theme resonates in three of the four selected papers~\cite{wang_enabling_2023,harrington_its_2022,jo_understanding_2023}. Moreover, the studies foreshadow benefits gained from aligning user and CUI expectations~\cite{yeh_how_2022,wang_enabling_2023} while informing about system capabilities~\cite{jo_understanding_2023} and making interactions transparent to increase trust~\cite{yeh_how_2022}. We hope our framework can guide future research and design of CUIs in a human-centred manner.


\section{Discussion \& Future Work}
Our research aimed to gain insights into the ethical caveats and design considerations faced in current and future CUI research and development. Motivated by dark pattern scholarship, particularly Mathur et al.~\cite{mathur2019} dark pattern characteristics, we conducted 27 semi-structured interviews and identified five themes
providing answers to our research question. Derived from these themes, we propose the CUI Expectation Cycle (CEC), a framework to guide future work in bridging CUI capabilities with users' expectations and goals. In this section, we return to our research question and discuss the different perspectives held by the interviewed cohorts. Lastly, we point toward some directions for future work.

\subsection{Revisiting Our Research Question}
With the main aim of this study to identify ethical caveats in CUI design, our findings answer our research question: Which ethical caveats should be considered when designing CUI interactions, and how should they be addressed? Our themes illustrate how the unique nature of CUIs cannot simply carry over expertise readily available for other domains. To fill this gap, the subordinate ethical caveats contain design considerations that, if disregarded, compromise user experience. These include the discoverability of features and possible interactions~\cite{branham_blind_users_2019,chen_technical_2018,wagner_alexa_2019, jain_2018, feine_2019}, often further limited to diverse and marginalised users~\cite{harrington_its_2022}, but also deceptive designs that lead to unrealistic expectation~\cite{luger_like_2016} and even unwanted interactions -- as seen in dark pattern literature~\cite{gray_2018_dark,mathur2019}.
Incorporating these themes, the CEC builds on established theory to provide practitioners and scholars with means to mitigate these effects.

\input{Tables/darkpatterns-per-theme}
\subsection{Perceptions Across Researchers, Practitioners, and Users}
Each developed theme considers the perspectives of each cohort, highlighting similarities and differences in the diverse code groups (the full codebook is included in the supplementary material of this paper). Although researchers and practitioners often shared similar expertise about the involved technologies, we noticed worries among users, illustrating a lack of the same technical understanding. Interestingly, this was further mirrored by similar views held among researchers and practitioners who -- in sum -- recognised a need to address users' concerns and develop CUIs that are more accessible and inclusive.

While all cohorts agree that CUIs should be more transparent about their capabilities and how data is handled, suggestions as to how this could be addressed differ. Researchers and frequent users pointed out that CUI design should reflect capabilities as anthropomorphic features overshadow limitations. Practitioners, noticing some relevance to announcing a CUI's artificial nature, argued for the benefits of CUIs offering companionship to lonely users. This is in line with previous work (e.g.~\cite{Voit_SmartSpeakers,Niess-Wozniak-Companion}, and mirrored by users admitting emotional bonds with their devices. However, both frequent users and practitioners expressed frustration regarding user experience. Interestingly, though, they are divided as to which group should take responsibility. While frequent users are frustrated about the lack of discoverability in hard-to-navigate interfaces, practitioners argue that users are not exhausting their CUI's technical potential. This disparity is one example of the gap between users and practitioners. A potential solution was presented across cohorts when discussing the need for improved onboarding for setting realistic expectations. Here, researcher and practitioner cohorts were aware of a need to address diverse and marginalised users better, while practitioners, in particular, described current technical limitations. A general need for guidelines is voiced by both researchers and practitioners. While the latter echoed this sentiment but missed standardised guidelines for CUIs, the former also noticed a need for better communication as contemporary research aims to fill this gap.


\subsection{Paving the Way for Future Work}
Our decision to listen to different voices allowed us to garner insights from three perspectives with partially contrasting incentives. Although related work has outlined similar concerns raised by our participants~\cite{doyle_mapping_2019,branham_blind_users_2019,yeh_how_2022,owens_exploring_2022,conca_present_2023}, our approach to connecting the discourse to unethical design and dark pattern literature has led to novel findings, resulting in further design considerations and bridging expectations between cohorts. Future work could aim to lift the tension between user and CUI and develop interfaces that meet their expectations. Thus, increasing user experience would be a success for the practitioners' work.

Our work can serve as a foundation for ethical design in CUI contexts. As the interview questions adopted Mathur et al.'s~\cite{mathur2019} dark pattern characteristics, we reviewed each theme as a potential host for dark patterns. For this, we follow Mildner et al.~\cite{mildner_about_2023} and draw from contemporary typologies that collectively describe over 80 types of dark patterns~\cite{mathur2019,mathur_what_2021,gray_2018_dark, gray_2020_asshole,conti_malicious_2010, bosch_2016_privacy, greenberg_2014_proxemic, gunawan_comparative_2021, zagal_dark_2013, brignull_deceptive_2022, mildner_about_2023}. Table~\ref{tab:darkpatterns_per_theme} allocates identified dark patterns, suggesting that each theme features nefarious opportunities for manipulative interactions. Notably, most dark pattern types have been described in GUI contexts. Introducing this body of work to CUI research offers valuable insights. However, our themes foreshadow additional patterns unique to CUI interactions, extending previously described instances in CUI contexts~\cite{owens_deceptive_2022,conca_present_2023}. As the technology is still in its relative infancy, understanding unethical practices early can present an important head start in protecting users. Future work could build on our findings to gain a better understanding of the underlying technologies of CUIs that could be exploited to deploy novel types of dark patterns.


\section{Limitations}

Although we were careful when designing and conducting this study, our work has limitations. The interviews were conducted using an online format, which, in individual cases, led to connectivity problems, prolonging some interviews. Although we ensured that all interviewees were asked the same questions, technological obstacles may have influenced participants' responses.


Sample size limitations should also be acknowledged, with a relatively small sample representing each cohort. That said, steps were taken to ensure cohorts were relatively balanced in size and gender and to ensure practitioners and researchers represented a relatively broad range of backgrounds and experiences. The size of each cohort sample is also common to qualitative research in the field, as similar qualitative studies estimate high saturation of codes within 7-12 interviews~\cite{guest_how_2006}. 
Nonetheless, this may pose limitations concerning the breadth of opinions expressed, which may not be representative across professions. Additionally, many practitioners have experience in research areas with multiple participants holding a Ph.D. For frequent users, we were mainly interested in people who use CUI devices weekly.  However, we acknowledge that we did not further inquire about the duration of participants' interactions with CUIs. Moreover, the fact that participants could afford devices may suggest a selection bias in this sample. While we aimed to sample frequent users from across the world, the size of this study hinders a representative sample. We used Prolific~\cite{prolific} to recruit matching participants, where we had to trust the platform and participants' self-evaluation to meet our criteria.


With regard to our themes, we recognise two limitations. First, participants drew heavily from their experience with voice-based interfaces. While we prompted each interviewee to consider both voice-based and text-based CUIs, replies from frequent users indicate insufficient experience with the latter. We were careful to frame our themes around all kinds of CUIs, but the lack of chatbot experience among the frequent user group may have introduced a bias toward voice-based systems. Second, not every design issue in our themes necessarily implies malicious intent. The current discourse accompanying dark patterns and malicious designs discusses whether the malevolent intents of practitioners are relevant or if any potential harm suffices to fulfil its criteria~\cite{grasl_dark_2021, gunawan_comparative_2021, obi_lets_2022}. Future work could address this gap to study potential links between intents and dark patterns to aid regulators and create guidelines for ethically aligned user interfaces.

\section{Conclusion}
Recent interest in HCI has raised awareness of unethical design in technologies. In this paper, we identified ethical caveats related to CUI technologies by conducting 27 interviews between researchers, practitioners, and frequent users of CUI systems. Based on our analysis, we identified five themes covering each group's perspectives and informing about exploitative designs' unethical consequences. In line with prior research, we noticed broken expectations between users and their devices' capabilities and learned about the underlying problems. We hope that the five ethical caveats, derived from our themes, can be used to support user-centred designs and create systems that are better aligned with service providers' intentions and users' expectations. To mitigate users' frustration and ensure more transparency and ethically aligned interactions, this work contributes a framework, the CUI Expectation Cycle, to connect users' expectations with CUIs' capabilities.

\begin{acks}
Special thanks go to Justin Edwards, whose great support has aided this work significantly throughout the project's course. The research of this work was partially supported by the Klaus Tschira Stiftung gGmbH. This work was conducted with the financial support of the Science Foundation Ireland Centre for Research Training in Digitally-Enhanced Reality (d-real) under Grant No. 18/CRT/6224.
This publication has emanated from research conducted with the financial support of Science Foundation Ireland under Grant number 12/RC/2289\_P2. 
For the purpose of Open Access, the authors have applied a CC BY public copyright licence to any Author Accepted Manuscript version arising from this submission. 
\end{acks}

\bibliographystyle{ACM-Reference-Format}
\bibliography{references.bib}

\newpage
\appendix
\clearpage
\section{Interview Questionnaire}\label{app:interview_questionnaire}

Here, we provide a Table~\ref{tab:mathur2019} including Mathur et al.'s~\cite{mathur2019} original characteristics as well as adapted questions. Below the table is the full script of all nine interview questions. We conditionally prompted interviewees to talk about graphical-based systems and voice-based systems as well as contemporary and future designs to get multi-faceted answers.

\begin{table}[h]
\begin{tabular}{p{0.2\linewidth}p{0.7\linewidth}}
\toprule
\multicolumn{2}{c}{\textbf{Original and adapted questions}} \\
\multicolumn{2}{c}{\textbf{from the dark pattern characteristics}} \\
\multicolumn{2}{c}{\textbf{by Mathur et al.~\cite{mathur2019}}} \\ \midrule
\renewcommand{\arraystretch}{1.4}
Characteristic  & Question (Originals are \textit{italic} whereas interview questions are not)\\ \midrule

\multirow{2}{*}{Asymmetry}          		& \textit{Does the user interface design impose unequal weights or burdens on the available choices presented to the user in the interface?} \\
											& Can you tell me about how the design of conversational user interfaces may place specific burdens on people in terms of understanding what choices are available to them during interactions? \\ \midrule

\multirow{2}{*}{Covert}		 	            & \textit{Is the effect of the user interface design choice hidden from the user?} \\
											& How clear are the consequences of making particular choices to people when using conversational user interfaces? \\ \midrule

\multirow{2}{*}{Deceptive}	           		&\textit{Does the user interface design induce false beliefs either through affirmative misstatements, misleading statements, or omissions?}\\
											& Can you tell me about how the design of conversational user interface may induce false beliefs? \\ \midrule

\multirow{2}{*}{Hides Info.}  		& \textit{Does the user interface obscure or delay the presentation of necessary information to the user? }\\
											& How could aspects of the design of conversational user interface obscure or delay important information from the user? \\ \midrule

\multirow{2}{*}{Restrictive}		        & \textit{Does the user interface restrict the set of choices available to users? } \\ 
											& Do you feel that all available choices are clear to people when using conversational user interfaces? \\

\bottomrule
\end{tabular}
\caption{This table lists the descriptive questions by Mathur et al. (2019)~\cite{mathur2019} for their dark pattern characteristics. It further contains the adapted interview questions.}
\Description[Dark Pattern Characteristics]{This table lists the introductory questions Mathur et al. (2019) gave for each dark pattern characteristic. The characteristics and their descriptive questions are as follows: 
1. Asymmetric: Does the user interface design impose unequal weights or burdens on the available choices presented to the user in the interface?

2. Covert: Is the effect of the user interface design choice hidden from the user?

3. Deceptive: Does the user interface restrict the set of choices available to users?

4. Hides Information: Does the user interface design induce false beliefs either through affirmative misstatements,
misleading statements, or omissions?

5. Restrictive: Does the user interface obscure or delay the presentation of necessary information
to the user?}
\label{tab:mathur2019}
\end{table}
\begin{enumerate}
    \item Could you please describe your own experience with CUIs?
    \item Can you tell me about how the design of conversational user interfaces may place specific burdens on people in terms of understanding what choices are available to them during interactions?
    \item How clear are the consequences of making particular choices to people when using conversational user interfaces?
    \item Do you feel that all available choices are clear to people when using conversational user interfaces?
    \item Can you tell me about how the design of conversational user interface may induce false beliefs? (these might include false beliefs about the system or in terms of the information it provides.)
    \item How could aspects of the design of conversational user interface obscure or delay important information from the user?
    \item Can you tell me about some inherent limitations of conversational user interfaces that are not always apparent to people who use them?
    \item Can you tell me about situational contexts that might make people more vulnerable to problematic design in CUI interactions?
    \item Can you tell me about specific groups of people who might be particularly vulnerable to problematic design in CUI interactions?
\end{enumerate}
\end{document}

%% file: Tables/particiant-table.tex
\begin{table*}[!ht]
\begin{tabular}{@{}llllll@{}}
\toprule
\multicolumn{6}{c}{\textbf{\textsc{Participant Table}}} \\ \midrule
ID & Age & Gender & Country of Residence & Occupation & Years of Experience \\ \midrule
\multicolumn{6}{c}{\textbf{Researcher}} \\ \midrule \rowcolor[HTML]{EFEFEF} 
R1 & 38 & male & USA & Professor & 16 \\
R2 & 27 & female & USA & Professor & 11 \\ \rowcolor[HTML]{EFEFEF} 
R3 & 35 & male & USA & PhD Candidate & 10 \\
R4 & 33 & female & South Korea & Assistant Professor & 5 \\ \rowcolor[HTML]{EFEFEF} 
R5 & 69 & male & United Kingdom & Professor & 48 \\
R6 & 32 & male & Germany & Postdoctoral Researcher & 12 \\ \rowcolor[HTML]{EFEFEF} 
R7 & 30 & female & Switzerland & PhD Candidate & 4 \\
R8 & 30 & not disclosed & Ireland & PhD Candidate & 5 \\ \midrule
\multicolumn{1}{r}{Mean} & = 36.75 & & & \multicolumn{1}{r}{Mean} & = 13.88\\
\multicolumn{1}{r}{SD} & = 13.46 & & & \multicolumn{1}{r}{SD} & = 14.4\\

\multicolumn{6}{c}{\textbf{Practitioner}} \\ \midrule \rowcolor[HTML]{EFEFEF} 
P1 & 56 & male & United Kingdom & Chief Science Officer & 15 \\
P2 & 49 & female & USA & Manager & 21 \\ \rowcolor[HTML]{EFEFEF} 
P3 & 40 & male & USA & Educational Research & 2 \\
P4 & 32 & male & UK & Tech. Consulting Manager & 6 \\ \rowcolor[HTML]{EFEFEF} 
P5 & 40 & female & Brazil & Research & 9 \\
P6 & 38 & female & USA & Conversation Designer & 10 \\ \rowcolor[HTML]{EFEFEF} 
P8 & 64 & female & USA & Communication Consultant & 35 \\
P9 & 42 & female & Ireland & Designer & 15 \\ \rowcolor[HTML]{EFEFEF} 
P10 & 58 & male & USA & Digital Business Executive & 25 \\ \midrule
\multicolumn{1}{r}{Mean} & = 46.56 & & & \multicolumn{1}{r}{Mean} & = 15.33\\
\multicolumn{1}{r}{SD} & = 10.74 & & & \multicolumn{1}{r}{SD} & = 10.28\\

\multicolumn{6}{c}{\textbf{Frequent User}} \\ \midrule
ID & Age & Gender & Country of Residence & Current Occupation & Highest Level of Education \\ \midrule \rowcolor[HTML]{EFEFEF} 
F1 & 26 & female & Mexico & ESL Teacher & Undergraduate \\ 
F2 & 26 & female & South Africa & Customer Service Rep & Postgraduate (or higher) \\ \rowcolor[HTML]{EFEFEF} 
F3 & 30 & male & South Africa & Media Analyst & Secondary/Vocational \\
F4 & 33 & male & South Africa & Construction & Secondary/Vocational \\ \rowcolor[HTML]{EFEFEF} 
F5 & 24 & male & South Africa & Student & Undergraduate \\
F6 & 25 & female & South Africa & Student & Postgraduate (or higher) \\ \rowcolor[HTML]{EFEFEF} 
F7 & 57 & male & United Kingdom & Retired & Postgraduate (or higher) \\
F8 & 45 & female & Italy & Remote Freelancer & Secondary/Vocational \\ \rowcolor[HTML]{EFEFEF} 
F9 & 22 & male & Poland & Frontend Developer & Undergraduate \\
F10 & 29 & female & Mexico & Healthcare Entrepreneur & Undergraduate \\ \midrule
\multicolumn{1}{r}{Mean} & = 31.7 & & & &\\
\multicolumn{1}{r}{SD} & = 11.02 & & & &\\
\bottomrule
\end{tabular}
\caption{This table presents our interview participants in three cohorts: researcher, practitioner, and frequent users. Notably, we did not have a participant P7. The associated participant cancelled the interview after IDs were already given to all participants. To avoid confusion, we retained our initial structure.}
\label{tab:participants}
\end{table*}

%% file: Tables/ethical-caveats.tex
\begin{table*}[!t]
\centering
\renewcommand{\arraystretch}{1.8}
\setlength{\tabcolsep}{0.02\linewidth}
\begin{tabular}{p{0.125\linewidth}p{0.38\linewidth}p{0.38\linewidth}}
\hline
\textbf{Theme} & \textbf{Ethical Caveat} & \textbf{Guiding Question} \\ \hline

\rowcolor[HTML]{EFEFEF} 
Building Trust and Guarding Privacy & Users feel vulnerable to use CUIs, posing a need for CUI developers to prioritise transparency and control over data handling. & Does the system/interaction provide accessible and transparent information about personal data with easy control thereof?\\ 

Guiding Through Interactions & Guidelines and frameworks need to educate developers to develop accessible CUIs that empower users with diverse technological literacy to confidently interact with available features. & Does the system/interaction adequately inform users about its technical capabilities to enable full utilisation of its features?\\ 

\rowcolor[HTML]{EFEFEF} 
Human-like Harmony & Anthropomorphic features should be implemented with care and in line with a CUI's capabilities to support intuitive and authentic interactions, preventing unrealistic expectations. & Does the system/interaction clarify the presence of anthropomorphic features to avoid misconceptions and unrealistic expectations?\\

Inclusivity and Diversity & The development and design of CUI interactions need to consider individual needs and characteristics of users, especially marginalised groups, ensuring equitable CUI interactions. & Does the system/interaction cater towards users with diverse needs, potentially through alternative interactions where otherwise inaccessible?\\ 

\rowcolor[HTML]{EFEFEF} 
Setting Expectations & CUI capabilities should avoid deceptive interactions and, instead, be transparent to users to prevent frustration and mistrust. & Does the system/interaction handle user prompts truthfully, clarifying the scope of its capabilities to provide realistic expectations?\\

\hline
\end{tabular}
\caption{This table summarises the five identified themes and design questions per ethical caveat.}
\label{tab:theme-descriptions}
\end{table*}

%% file: Tables/darkpatterns-per-theme.tex
\begin{table*}[t!]
\resizebox{1\textwidth}{!}{%
\centering
\begin{tabular}{p{0.005\linewidth}p{0.2\linewidth}p{0.2\linewidth}p{0.2\linewidth}p{0.2\linewidth}p{0.2\linewidth}}
\toprule
\multicolumn{6}{c}{\textbf{Themes}} \\ \midrule
\multirow{17}{*}{\rotatebox[origin=c]{90}{\small\textbf{Dark Patterns}}}

& \multicolumn{1}{l}{\small{\begin{tabular}[c]{@{}c@{}}\textbf{Building Trust and} \\ \textbf{Guarding Privacy}\end{tabular}}}

& \multicolumn{1}{l}{\small{\begin{tabular}[c]{@{}c@{}}\textbf{Guiding Through} \\ \textbf{Interactions}\end{tabular}}} 

& \multicolumn{1}{l}{\small{\begin{tabular}[c]{@{}c@{}}\textbf{Human-like} \\ \textbf{Harmony}\end{tabular}}} 

& \multicolumn{1}{l}{\small{\begin{tabular}[c]{@{}c@{}}\textbf{Inclusivity} \\ \textbf{and Diversity}\end{tabular}}} 

& \multicolumn{1}{l}{\small{\begin{tabular}[c]{@{}c@{}}\textbf{Setting} \\ \textbf{Expectations}\end{tabular}}}\\ \midrule

& \small{\begin{tabular}[t]{|l}
    · Bad Defaults~\cite{bosch_2016_privacy}                                \\
    \begin{tabular}[l]{@{}l@{}}· Captive                                    \\
    \hspace{5pt}Audience~\cite{greenberg_2014_proxemic}\end{tabular}        \\
    · Disguised Ads~\cite{brignull_deceptive_2022}                          \\
    \begin{tabular}[l]{@{}l@{}}· Disguised Data                             \\
    \hspace{5pt}Collection~\cite{greenberg_2014_proxemic}\end{tabular}      \\
    \begin{tabular}[l]{@{}l@{}}· Hidden Legalese                            \\
    \hspace{5pt}Stipulations~\cite{bosch_2016_privacy}\end{tabular}         \\
    \begin{tabular}[l]{@{}l@{}}· Hides                                      \\
    \hspace{5pt}Information~\cite{mathur2019,mathur_what_2021}\end{tabular} \\
    \begin{tabular}[l]{@{}l@{}}· Making Personal                            \\
    \hspace{5pt}Information                                                 \\
    \hspace{5pt}Public~\cite{greenberg_2014_proxemic}\end{tabular}          \\
    \begin{tabular}[l]{@{}l@{}}· Privacy                                    \\
    \hspace{5pt}Zuckering~\cite{brignull_deceptive_2022}\end{tabular}       \\
    \end{tabular}} 

& \small{\begin{tabular}[t]{l}
    · Confusion~\cite{conti_malicious_2010}                                 \\
    \begin{tabular}[l]{@{}l@{}}· Exploiting                                 \\
    \hspace{5pt}Errors~\cite{conti_malicious_2010}\end{tabular}             \\
    \begin{tabular}[l]{@{}l@{}}· Manipulating                               \\
    \hspace{5pt}Navigation~\cite{conti_malicious_2010}\end{tabular}         \\
    · Misrepresenting~\cite{gray_2020_asshole}                              \\
    · Obfuscation~\cite{conti_malicious_2010}                               \\
    · Obstruction~\cite{gray_2018_dark}                                     \\
    · Roach Motel~\cite{brignull_deceptive_2022}                            \\
    · Trick~\cite{conti_malicious_2010}                                     \\
\end{tabular}}

& \multicolumn{1}{c}{\small{\begin{tabular}[t]{l}
    · Deceptive~\cite{mathur2019, mathur_what_2021}                         \\
    \begin{tabular}[l]{@{}l@{}}· Hides                                      \\
    \hspace{5pt}Information~\cite{mathur2019,mathur_what_2021}\end{tabular} \\
    · Misrepresenting~\cite{gray_2020_asshole}                              \\
    · Sneaking~\cite{gray_2018_dark}                                        \\
    · Trick~\cite{conti_malicious_2010}                                     \\
\end{tabular}}} 

& \small{\begin{tabular}[t]{l}
    · Covert~\cite{mathur2019,mathur_what_2021}                             \\
    \begin{tabular}[l]{@{}l@{}}· Disparate                                  \\
    \hspace{5pt}Treatment~\cite{mathur_what_2021}\end{tabular}              \\
    \begin{tabular}[l]{@{}l@{}}· Making Personal                            \\
    \hspace{5pt}Information                                                 \\
    \hspace{5pt}Public~\cite{greenberg_2014_proxemic}\end{tabular}          \\
    \begin{tabular}[l]{@{}l@{}}· Social Network                             \\
    \hspace{5pt}Of Proxemic                                                 \\
    \hspace{5pt}Contacts Or                                                 \\
    \hspace{5pt}Unintended                                                  \\
    \hspace{5pt}Relationships~\cite{greenberg_2014_proxemic}\end{tabular}   \\
\end{tabular}} 

& \small{\begin{tabular}[t]{l}
    · Asymmetry~\cite{mathur_what_2021}                                     \\
    \begin{tabular}[l]{@{}l@{}}· Automating                                 \\
    \hspace{5pt}The User~\cite{gray_2020_asshole}\end{tabular}              \\
    · Bait \& Switch~\cite{bosch_2016_privacy, greenberg_2014_proxemic}     \\
    · Forced Actions~\cite{gray_2018_dark}                                  \\
    · Forced Work~\cite{conti_malicious_2010}                               \\
    \begin{tabular}[l]{@{}l@{}}· Hidden Legalese                            \\
    \hspace{5pt}Stipulations~\cite{bosch_2016_privacy}\end{tabular}         \\
    \begin{tabular}[l]{@{}l@{}}· Interface                                  \\
    \hspace{5pt}Interference~\cite{gray_2018_dark}\end{tabular}             \\
    · Obfuscation~\cite{conti_malicious_2010}                               \\
    · Obstruction~\cite{gray_2018_dark}                                     \\
    \begin{tabular}[l]{@{}l@{}}· Restricting                                \\
    \hspace{5pt}Functionalities~\cite{conti_malicious_2010}\end{tabular}    \\
\end{tabular}}

\\ \bottomrule
\end{tabular}
}
\caption{This table summarises previously captured dark patterns within each theme.}
\label{tab:darkpatterns_per_theme}
\end{table*}